%% file: main.tex
\documentclass[a4paper, amsfonts, amssymb, amsmath, reprint, showkeys, nofootinbib, twoside]{revtex4-1}
\usepackage[english]{babel}
\usepackage[utf8]{inputenc}
\usepackage[colorinlistoftodos, color=green!40, prependcaption]{todonotes}
\input{preamble}
\usepackage[pdftex, pdftitle={Article}, pdfauthor={Author}]{hyperref} % For hyperlinks in the PDF
\usepackage{amsmath, bm}

\begin{document}
%\title{Spin Signatures in the Fusion of Majorana Vortices in a Planar Josephson Junction}
\title{Spin-Dependent Signatures of Majorana Vortex Fusion within Planar Josephson Junctions}

\author{Krishnan Ganesh}
    \email[Correspondence email address: ]{k.ganesh21@imperial.ac.uk}% Your name
    \affiliation{Blackett Laboratory, Imperial College London, London SW7 2AZ, United Kingdom}
\author{Derek K. K. Lee}
\affiliation{Blackett Laboratory, Imperial College London, London SW7 2AZ, United Kingdom}
\author{Jiannis K. Pachos}
\affiliation{School of Physics and Astronomy, University of Leeds, Leeds LS2 9JT, United Kingdom}

\date{\today} % Leave empty to omit a date

\begin{abstract}

We investigate the magnetic characteristics and tunnelling signatures of a planar Josephson junction with Rashba spin-orbit coupling during the fusion of two Majorana vortices. By employing the topological phase diagram and conducting tight-binding simulations of the proposed device, we demonstrate that this fusion process induces a parity-dependent magnetic moment aligned with the junction axis. We further propose a method to probe the spin properties of the fusing Majorana zero modes through spin-resolved Andreev conductance measurements at the junction endpoints. To support our findings, we derive a low-energy effective Hamiltonian that provides a detailed microscopic description of the numerically observed phenomena. Our analysis enables the detection of Majorana fusion outcome from accessible spin current measurements, thus paving the way for future experimental verification and potential applications in topological quantum computation.
\end{abstract}


\maketitle

\input{sections/section01.tex} 
\input{sections/section02.tex}
\input{sections/section03.tex}
\input{sections/section04.tex}
\input{sections/section05.tex}
\input{sections/section06.tex}

\input{sections/acknowledgements.tex}

\bibliographystyle{unsrt}
\bibliography{references.bib}

\appendix*
\input{sections/appendix1.tex}

\end{document}

%% file: preamble.tex
\usepackage{amsthm}
\usepackage{mathtools}
\usepackage{physics}
\usepackage{xcolor}
\usepackage{graphicx}
\usepackage[left=23mm,right=13mm,top=35mm,columnsep=15pt]{geometry} 
\usepackage{adjustbox}
\usepackage{placeins}
\usepackage[T1]{fontenc}
\usepackage{lipsum}
\usepackage{csquotes}
\usepackage{xfrac}
\usepackage{braket}

%% file: sections/section01.tex
\section{Introduction} \label{sec:introduction}
\begin{figure}
    \centering
    \includegraphics[width= 0.45\textwidth]{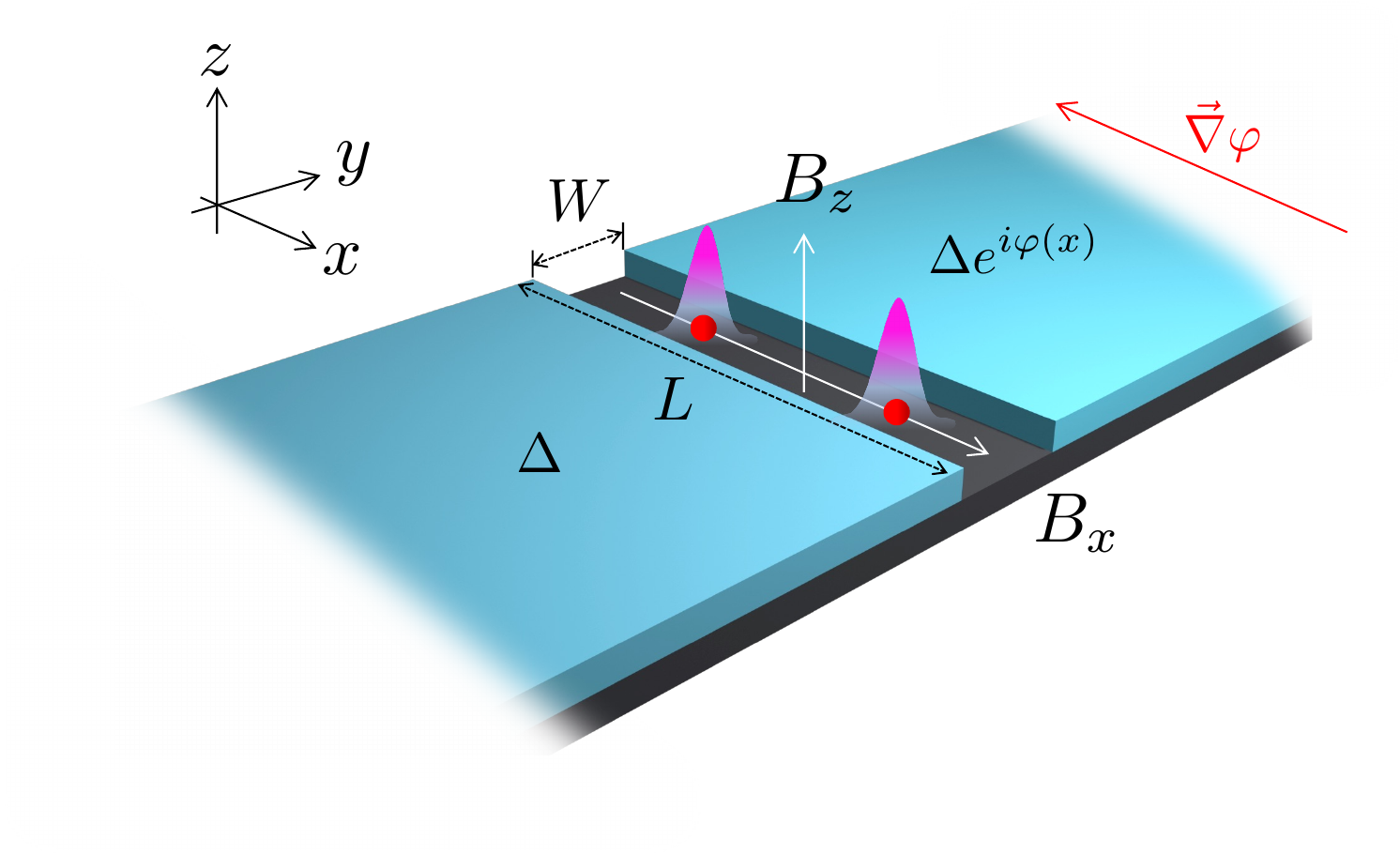}
    \caption{
Schematic illustration of a planar Josephson junction hosting Majorana zero modes at topological domain walls. The junction, with width $W$ and length $L$, comprises a two-dimensional electron gas (2DEG) with strong Rashba spin-orbit coupling (grey), contacted by two $s$-wave superconductors (blue) with pairing potential $\Delta$. An in-plane magnetic field $B_x$ induces the topological phase, while the out-of-plane magnetic field component $B_z$ generates a Josephson phase gradient $\nabla \varphi(x)$. This gradient results in the formation of topological domain walls (red dots), each of which binds a Majorana zero mode, represented by the purple wavefunctions.
}
\label{fig:Rashba_junction}
\end{figure}

One of the defining features of non-Abelian anyons is their ability to fuse into different particle types. For example, Majorana zero modes (MZMs), denoted as $
{\boldsymbol \gamma}$, can fuse to form either the vacuum state, denoted as ${\boldsymbol 1}$, or a fermionic quasiparticle, denoted as ${\boldsymbol \psi}$. This fusion process is mathematically captured by the fusion rule \cite{beenakker2020search, nayak2008non}
\begin{equation}\label{eq: Ising anyon Fusion rule} {\boldsymbol \gamma} \times {\boldsymbol \gamma} = {\boldsymbol 1} + {\boldsymbol \psi}. \end{equation}
Over the past decade, MZMs have gained significant interest since they have been theoretically predicted to emerge in topological superconductors (TSCs) \cite{kitaev2001unpaired, fendley2007edge}. These MZMs are typically localised at topological defects, such as the ends of a topological superconductor or within Abrikosov vortices in the bulk \cite{cheng2010tunneling, ivanov2001non, machida2023searching}. The fusion of two MZMs can result in either a fully paired state with an even number of particles, thus corresponding to the vacuum ${\boldsymbol 1}$, or a state with an unpaired quasiparticle, ${\boldsymbol \psi}$. Distinguishing between these two fusion outcomes will enable the demonstration of non-Abelian statistics in topological superconductors and thus open the way for realising topological qubits.\\ \linebreak
Numerous proposals for realising topological superconductors have concentrated on nanowires with strong spin-orbit coupling, which are proximitized to $s$-wave superconductors \cite{sau2021topological, frolov2020topological, mourik2012signatures, fu2008superconducting, vaitiekenas2020flux, yazdani2023hunting}. The literature also offers a wide range of possible methods for measuring the fusion channel of two Majorana zero modes in these systems. These approaches include coupling the nanowire to a quantum dot \cite{bai2023probing, gharavi2016readout, steiner2020readout, PhysRevB.101.235441}, embedding the nanowire within a Josephson flux qubit \cite{vijay2016teleportation}, and integrating the wire into an Aharonov-Bohm interferometer \cite{plugge2017majorana, aghaee2024interferometric}. However, to date there have been very few experimental implementations of these protocols due to challenges in identifying the topological phase in these devices. Thus, there is an active interest in investigating  alternative devices and searching for new signatures of the topological properties of Majorana zero modes that can conclusively demonstrate their non-Abelian character. 
\\ \linebreak
In this letter, we focus on a recent proposal to realise topological superconductivity within a planar Josephson junction \cite{pientka2017topological}. This system consists of a two-dimensional electron gas (2DEG) with strong Rashba spin-orbit coupling, contacted by two $s$-wave superconductors and subjected to a magnetic field $\vec{B}$, as shown in Figure \ref{fig:Rashba_junction}. This device exhibits extensive regions of topological superconductivity as the phase bias $\varphi$ and the in-plane magnetic field $B_{x}$ are varied \cite{pientka2017topological, setiawan2019topological, ren2019topological, fornieri2019evidence, dartiailh2021phase, pekerten2024microwave, melo2023greedy}. Further theoretical studies demonstrated that an out-of-plane magnetic field can generate Josephson vortices which induce topological domain walls that exponentially localise Majorana zero modes \cite{stern2019fractional}. 
%These domain walls can be manipulated by adjusting the phase bias between the superconducting leads, enabling the braiding of Majorana fermions in a three-terminal geometry. 
Our focus here is on the challenge of reading out the fusion channel of the Majoranas. In contrast to previous studies of Majorana fusion, such as Ref. \cite{gharavi2016readout, steiner2020readout}, which focus on a charge based signature, we investigate the potential to distinguish between the even and odd parity states of two topological domain walls by probing the magnetisation of the Josephson junction. \\ \linebreak
In particular, we investigate the spin properties and tunable coupling of Majorana zero modes in a planar Josephson junction with strong Rashba spin-orbit coupling. We show that the in-plane Zeeman field enables control over the separation and coupling of Majorana modes, influencing their energy splitting. By deriving an effective Hamiltonian, we describe their localization near topological domain walls and analyze how spin-dependent Andreev conductance can be used to experimentally detect the spin characteristics of the fusing Majorana zero modes. 

%% file: sections/section02.tex
\section{Josephson Vortices and Topological Domain Walls} \label{sec: Model}
To investigate the behaviour of the planar Josephson junction depicted in Fig. \ref{fig:Rashba_junction}, we first consider the case where the out-of-plane magnetic field $B_{z} = 0$. Hence, the applied magnetic field is purely in the $x$-direction and decays exponentially into the superconducting leads due to Meissner screening. Assuming the London penetration depth of the superconducting leads is small compared to the width of the junction $W$ and the coherence length of the superconductor, we can approximate the in-plane magnetic field as
\begin{equation}
B_{x}(y) = B \ \vartheta(W/2 - |y|),
\end{equation}
where $\vartheta(y)$ is the Heaviside step function.  We also assume the system is in the quasi-2D electron gas regime, so orbital effects of the in-plane magnetic field are negligible. The Hamiltonian written in the Nambu basis  $\Psi(\bm{r}) = \left( c_{\uparrow}(\bm{r}) , c_{\downarrow}(\bm{r}) , c^{\dagger}_{\downarrow}(\bm{r}) , -c^{\dagger}_{\uparrow}(\bm{r})\right)^{T}$ is 
\begin{equation}
H = \frac{1}{2}\int d^{2}r \Psi^{\dagger}(\bm{r})\mathcal{H}\Psi(\bm{r}),    
\end{equation}
The Bogoliubov-de-Gennes Hamiltonian for the system reads ($\hbar = 1$):
\begin{multline}\label{eqn: Bogoliubov de Gennes Hamiltonian for system}
\mathcal{H} = \left(\frac{-\bm{\nabla}^{2}}{2m} - \mu\right)\sigma_{0}\tau_{z} + \alpha \left( \bm{k} \times {\boldsymbol \sigma}\right)\cdot \hat{z}\tau_{z} \\ + \Delta(y) \sigma_{0}\tau_{+} + \Delta^{*}(y)\sigma_{0}\tau_{-} + E_{Z}(y)\sigma_{x}\tau_{0},
\end{multline}
where $m$ is the effective electron mass, $\mu$ is the chemical potential, $\alpha$ is the Rashba spin-orbit coupling energy and $\Delta(y)$ is the superconducting pair potential, which is approximated by
\begin{equation}
\Delta(y) = \vartheta(y - W)\Delta_{0}e^{i\varphi} + \vartheta(- y) \Delta_{0}.
\end{equation}
The Pauli matrices $\sigma_{i}$ and $\tau_{i}$ act on spin and particle-hole space respectively, so the notation $\sigma_{i}\tau_{j}$ is shorthand for the Kronecker product $\sigma_{i} \otimes \tau_{j}$. We have also defined the raising and lowering operators $\tau_{\pm} = (\tau_{x} \pm i \tau_{y})/2$.
The Hamiltonian $\mathcal{H}$ anticommutes with the particle-hole symmetry operator, which is
 $P = \sigma_{y}\tau_{y}K$,
in the Nambu basis where $K$ stands for complex conjugation. We perform our numerical investigation on a square lattice with lattice constant $a$, hopping parameter $t = 1/2ma^{2}$ and the tight-binding Hamiltonian given in Equation \ref{eqn: tight binding Hamiltonian}. All energies are given in units of the hopping parameter $t$.  \\ \linebreak
\begin{figure*}
    \centering
    \includegraphics[width=0.854\linewidth]{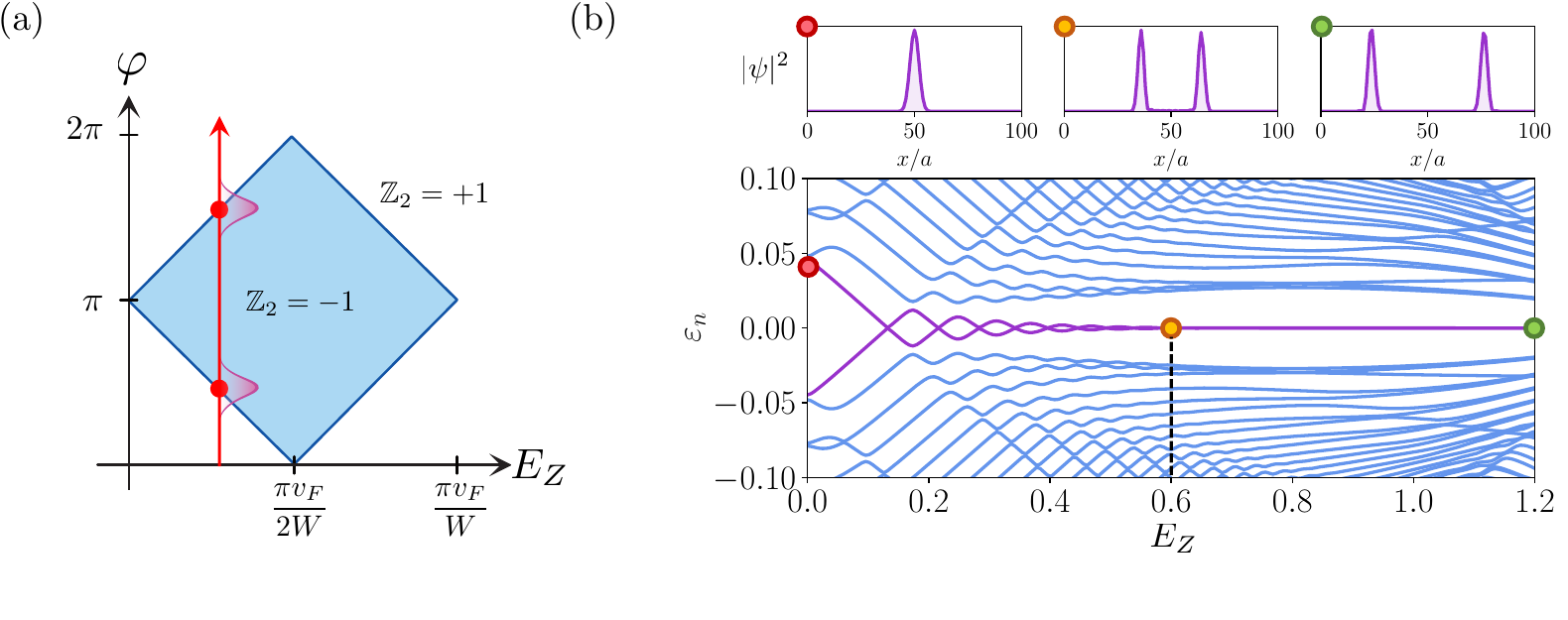}
    \caption{\textbf{(a)} Topological phase diagram for planar Josephson junction as a function Josephson phase difference $\varphi$ and in-plane Zeeman coupling $E_{Z}$ in the absence of an out-of-plane magnetic flux. The blue region indicates topologically non-trivial phase. In the presence of an out-of-plane flux $\Phi  = h/2e$, a Josephson vortex phase texture (red arrow) nucleates topological domain walls (red discs), whose separation depends on $E_{Z}$. \textbf{(b)} Top: Probability densities, $|\psi|^{2}$, of the lowest energy eigenstates along the normal region of the junction, shown for in-plane Zeeman field strengths $E_{Z} = 0$, $0.6$, and $1.2$. Bottom: Single-particle energy spectrum as a function of $E_{Z}$ for $\Phi = \Phi_{S}$. As $E_{Z}$ decreases, the two degenerate Majorana zero modes (purple) hybridize and split in energy. Coloured dots correspond to the eigenstates plotted in the top panels. The tight binding parameters used are: 
    $\mu/t = 1.8$, $\alpha/t = 0.3$, $\Delta_{0}/t = 0.8$, $W = a$, $L = 100a$ and the width of the superconducting contacts is $W_{SC} = 20a$.}
    \label{fig:phase diagram and wavefunctions_3}
\end{figure*}
As described in Ref.\cite{pientka2017topological, stern2019fractional} and further detailed in Appendix \ref{appendix: topological phase transitions}, this model exhibits topological phase transitions at Josephson phase differences of
\begin{equation}
\varphi_{\pm} = \pi \pm \frac{2E_{Z}W}{v_{F}},
\end{equation}
where $v_{F}$ is the Fermi velocity. As shown in Figure \ref{fig:phase diagram and wavefunctions_3}(a), this gives rise to the diamond regions of topological superconductivity as $\varphi$ and $E_{Z}$ are varied.\\ \linebreak
Let us now consider the presence of a small out-of-plane field, the Josephson phase difference winds linearly in the $x$-direction \cite{vecpotentialandscreening}\cite{gross2016applied}
\begin{equation}
    \varphi(x) = \frac{2\pi \Phi}{\Phi_{S} L}x - \theta,
\end{equation}
where $\Phi = B_{z}LW$ is the flux through the junction, $\Phi_{S} = h/2e$ and $\theta$ is a global phase shift, which generates topological domain walls at positions where $\varphi(x) = \varphi_{\pm}$, each binding a single Majorana zero mode. 

%% file: sections/section03.tex
\section{Tunable Coupling Between Majorana Zero Modes} \label{sec: Tunable Coupling between Majorana Zero modes}
The in-plane Zeeman coupling provides a useful experimental control for fusing topological domain walls. As illustrated in the top panel of Figure \ref{fig:phase diagram and wavefunctions_3}(b), varying $E_{Z}$ controls the separation between the localised Majorana zero mode wavefunctions. When $E_{Z} \approx 0$, the Majorana zero modes strongly couple, forming a localised fermionic mode. This behaviour is also reflected in the low-energy spectrum of the Josephson junction as shown in bottom panel of Figure \ref{fig:phase diagram and wavefunctions_3}(b). As the Zeeman field is reduced, the energy splitting between the Majorana zero modes grow exponentially, accompanied by oscillations.\\ \linebreak 
These features can be heuristically understood by considering the case of a single Josephson vortex phase distribution, $\varphi(x) = 2\pi x/L$, with both $E_{Z}$ and the spin-orbit coupling $\alpha$ set to zero. In this scenario, a discrete quasiparticle spectrum $\varepsilon_{n}$ emerges, which is approximately spin-degenerate at $E_{Z} = 0$, up to a small Zeeman field in the $z$-direction which we neglect \cite{PhysRevB.105.094516}. As the in-plane Zeeman coupling is increased, the spin-degenerate levels split resulting in a spectrum $E_{n,\sigma}(E_{Z}) = \varepsilon_{n} +\sigma E_{Z}$, where $\sigma = \pm 1$ labels the $\sigma_x$ eigenvalue of the state. Level crossings occur at zero energy for Zeeman coupling values $E_{z, n} = \varepsilon_{n}$. Upon increasing $\alpha$, level repulsions occur between $\sigma = +1$ states and $\sigma = -1$ states, which eventually separates two states, shown in purple in Figure \ref{fig:phase diagram and wavefunctions_3}(b) (bottom), from the rest of the spectrum, shown in blue. Therefore, the oscillations in the energy splitting appear to originate from the discrete level structure of the Josephson vortex in the absence of Zeeman splitting and spin-orbit coupling.

%% file: sections/section04.tex
\section{Spin Characteristics}\label{section: Spin characteristics}

For $\Phi = \Phi_{S}$, close to the coordinates of the topological domain wall the Majorana zero mode wavefunctions can be approximated by 
\begin{equation}\label{eq: approximate Majorana wavefunction form}
    \psi_{\pm}(x) \sim \exp(\frac{-1}{\xi}|x - x_{\pm}|),
\end{equation}
where $x_{\pm} = L\varphi_{\pm}/2\pi$ and $\xi$ is the localisation length, which is related to the bulk topological gap. Therefore, the tunnel splitting between the zero modes can be estimated to be 
\begin{equation}\label{eq: approximate energy level splitting}
    \delta E \sim \exp(\frac{-|x_{+} - x_{-}|}{\xi}) = \exp(-\frac{4LE_{Z}W}{\pi \xi v_{F}}).
\end{equation}
The coupling between Majorana zero modes $\hat{\gamma}_{1}$ and $\hat{\gamma}_{2}$ introduces the term 
\begin{equation}
    \hat{H}_{tun.} = \frac{i\delta E}{2} \hat{\gamma}_{1}\hat{\gamma}_{2}
\end{equation}
in the Hamiltonian. Given that $\delta E$ has an exponential dependence on the in-plane field $B_{x}$, a clear physical observable that depends on the joint parity of two Majorana zero modes is the \textit{magnetisation} of the Andreev bound states, which we may define as 
\begin{equation}\label{eq: magnetisation of the junction}
    \bm{m} = \Bra{\Omega}\frac{\partial \hat{H}_{tun.}}{\partial \bm{B}}\Ket{\Omega},
\end{equation}
where $\ket{\Omega}$ is a many-body ground state of the superconductor. Since the total Hamiltonian commutes with the parity operator $\hat{\mathcal{P}} = i\hat{\gamma}_{1}\hat{\gamma}_{2}$, $\ket{\Omega}$ must be an eigenstate of $\hat{\mathcal{P}}$ with eigenvalue $+1$ or $-1$. Using the Feynman-Hellmann theorem, the magnetisation can be expressed as 
\begin{equation}
\bm{m}_{\pm} = \pm \frac{\partial(\delta E)}{\partial \bm{B}}
\end{equation}
where $\pm$ is the parity eigenvalue. Since the out-of-plane magnetic field in our model is very small, the magnetisation should predominantly point in the $\pm x$ directions depending on the parity of the Majorana zero modes.
Furthermore, we numerically calculate the spin of the many-body state as a function of the in-plane Zeeman coupling $E_{Z}$. The spin-operator in second quantised form is given by 
\begin{equation}
    \hat{S}^{\mu}_{i} = \sum_{\sigma , \sigma'} \hat{c}^{\dagger}_{i\sigma}S^{\mu}_{\sigma\sigma'}\hat{c}_{i\sigma'}
\end{equation}
where $\mu \in \{x , y , z\}$, $i$ is a site index and $\sigma \in {\uparrow , \downarrow}$. We calculate the expectation value of this operator in the even and odd parity ground states $\ket{\Omega_{+}}$ and $\ket{\Omega_{-}}$, respectively. The diagonalised many-body Hamiltonian reads
\begin{equation}
    \hat{H} = \sum_{n = 0}^{2N-1} \varepsilon_{n}\hat{b}^{\dagger}_{n}\hat{b}_{n}
\end{equation}
where $\varepsilon_{n} > 0$ and $N$ is the number of lattice sites in the system. The even-parity ground state $\ket{\Omega_{+}}$ is annihilated by all the Bogoliubov destruction operators $\hat{b}_{n}$. At zero temperature, the odd-parity state will have the lowest energy quasiparticle occupied:
\begin{equation}
    \ket{\Omega_{-}} = \hat{b}^{\dagger}_{0}\ket{\Omega_{+}}.
\end{equation} 
The expectation values $\langle \hat{S}^{\mu}_{i}\rangle_{\pm} = \bra{\Omega_{\pm}}\hat{S}^{\mu}_{i}\ket{\Omega_{\pm}}$ are calculated using the inverse Bogoliubov transformation
\begin{align}
    \hat{c}_{i\uparrow} &= \sum_{n = 0}^{2N - 1}u^{n}_{i\uparrow}\hat{b}_{n} - (v^{n}_{i \uparrow})^{*}\hat{b}^{\dagger}_{n} \\
    \hat{c}_{i\downarrow} &= \sum_{n = 0}^{2N - 1} u^{n}_{i\downarrow}\hat{b}_{n} + (v^{n}_{i \downarrow})^{*}\hat{b}^{\dagger}_{n} \\
    \hat{c}^{\dagger}_{i\downarrow} &= \sum_{n = 0}^{2N - 1} (u^{n}_{i\downarrow})^{*}\hat{b}^{\dagger}_{n} + v^{n}_{i \downarrow}\hat{b}_{n} \\
    \hat{c}^{\dagger}_{i\uparrow} &= \sum_{n = 0}^{2N - 1}(u^{n}_{i\uparrow})^{*}\hat{b}^{\dagger}_{n} - v^{n}_{i \uparrow}\hat{b}_{n},
\end{align}
where the eigenvector corresponding to the $n^{th}$ positive energy state is given by $\left(u^{n}_{i\uparrow} , u^{n}_{i\downarrow} , v^{n}_{i\downarrow}  , v^{n}_{i\uparrow} \right)^{T}$. Using the property $\hat{b}_{n}\ket{\Omega_{+}} = 0 \ \forall n$, the expectation values are 
\begin{align}
    \langle\hat{S}^{\mu}_{x}\rangle_{+} &= -\sum_{n = 0}^{2N - 1}\left( v_{i\uparrow}^{n}v^{n *}_{i\downarrow} +v_{i\downarrow}^{n}v^{n *}_{i\uparrow}\right) \\
    \langle\hat{S}^{\mu}_{y}\rangle_{+} &=  -\sum_{n = 0}^{2N - 1}\left( -iv_{i\uparrow}^{n}v^{n *}_{i\downarrow} + iv_{i\downarrow}^{n}v^{n *}_{i\uparrow}\right) \\
    \langle\hat{S}^{\mu}_{z}\rangle_{+} &=  \sum_{n = 0}^{2N - 1}\left( v_{i\uparrow}^{n}v^{n *}_{i\uparrow} - v_{i\downarrow}^{n}v^{n *}_{i\downarrow}\right).
\end{align} 
The expectation values in the state $\ket{\Omega_{-}}$ are given by 
\begin{align}
    \langle\hat{S}^{\mu}_{x}\rangle_{-} &= \langle\hat{S}^{\mu}_{x}\rangle_{+} + \sum_{\sigma\sigma'}u_{i\sigma}^{0*}S^{x}_{\sigma\sigma'}u_{i\sigma'}^{0} + v_{i\sigma}^{0}S^{x}_{\sigma\sigma'}v_{i\sigma'}^{0*} \\
    \langle\hat{S}^{\mu}_{y}\rangle_{-} &= \langle\hat{S}^{\mu}_{y}\rangle_{+} + \sum_{\sigma\sigma'}u_{i\sigma}^{0*}S^{y}_{\sigma\sigma'}u_{i\sigma'}^{0} + v_{i\sigma}^{0}S^{y}_{\sigma\sigma'}v_{i\sigma'}^{0*} \\
    \langle\hat{S}^{\mu}_{z}\rangle_{-} &= \langle\hat{S}^{\mu}_{z}\rangle_{+} + \sum_{\sigma\sigma'}u_{i\sigma}^{0*}S^{z}_{\sigma\sigma'}u_{i\sigma'}^{0} - v_{i\sigma}^{0}S^{z}_{\sigma\sigma'}v_{i\sigma'}^{0*}.
\end{align}
\begin{figure}
    \centering    \includegraphics[width=1.0\linewidth]{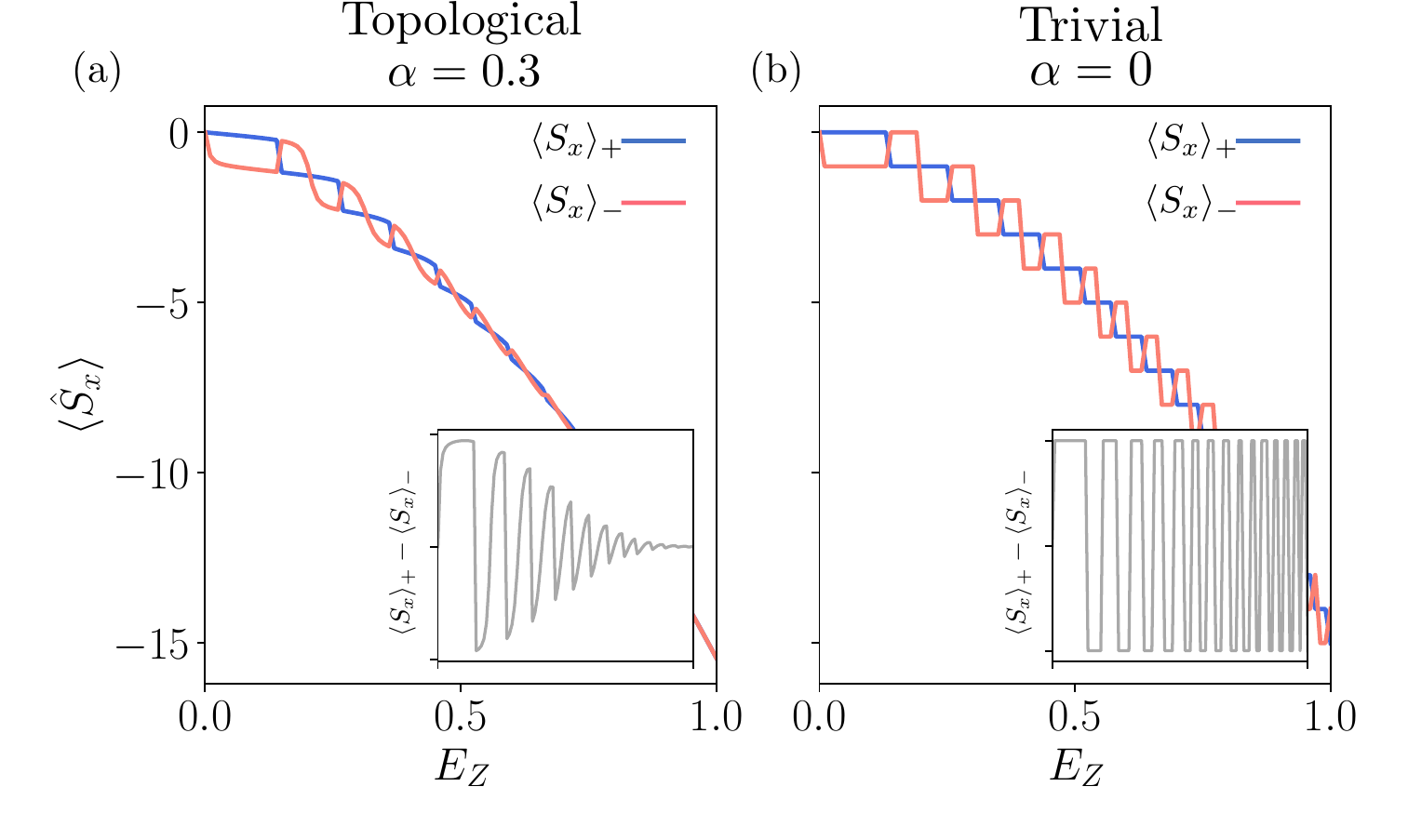}
    \caption{Absolute values of the spin expectation values for the even (blue) and odd (red) parity sectors. \textbf{(a)} Topological phase with $\alpha/t = 0.3$. \textbf{(b)} Trivial phase with $\alpha = 0$. \textbf{Insets:} Difference in the many-body spin expectation value along the $x$-axis between the even parity state $\langle S_{x} \rangle_{+}$ and the odd parity state $\langle S_{x} \rangle_{-}$ as a function of the in-plane Zeeman coupling $E_{Z}$.   The expectation values of the $y$ and $z$ spin components, $\langle S_{y}\rangle_{\pm}$ and $\langle S_{z}\rangle_{\pm}$, are approximately zero. The tight-binding parameters used are identical to those in Figure \ref{fig:phase diagram and wavefunctions_3}(b).}
    \label{fig: spin expectation value}
    
\end{figure} 
In Figure \ref{fig: spin expectation value} we plot the value of $\langle \hat{S}_{x}\rangle = \sum_{i}\langle \hat{S}^{x}_{i} \rangle$ for the even and odd parity states as well their absolute values in the insets. We do not plot $\langle\hat{S}_{y}\rangle$ and $\langle \hat{S}_{z}\rangle$ because they were found to be zero. In Figure \ref{fig: spin expectation value}(a) a clear difference in $\langle S_{x} \rangle$ for small \textit{but finite} $E_{Z}$ is observed, whereas for large $E_{Z}$ the states of different parity are indistinguishable. We also repeated the calculation for a trivial Josephson junction in Figure \ref{fig: spin expectation value}(b) by setting $\alpha = 0$. In contrast to the case with topological domain walls, in Figure \ref{fig: spin expectation value}(b) there is always a clear difference in the values of $\langle S_{x}\rangle_{\pm}$. At exactly $E_{Z} = 0$ however, the difference between $\langle S_{x}\rangle _{+}$ and $\langle S_{x} \rangle_{-}$ is zero for both the topological and trivial junction because the bound states are spin-degenerate. Hence, the spin signature of Majorana fusion can only be observed as $E_{Z} \to 0$.
% Furthermore, level crossings at $\varepsilon = 0$ coincide with switches in the value of $\langle S_{x}\rangle$ for the states $\ket{\Omega_{+}}$ and $\ket{\Omega_{-}}$. This is because level crossings at $\varepsilon = 0$ correspond to fermion-parity switches, where the energy associated with exciting a quasiparticle becomes negative, making the odd-parity state energetically favourable.

%% file: sections/section05.tex
\section{Spin dependent Andreev conductance} \label{sec: spin dependent Andreev conductance}
\begin{figure}
    \centering
    \includegraphics[width=1.0\linewidth]{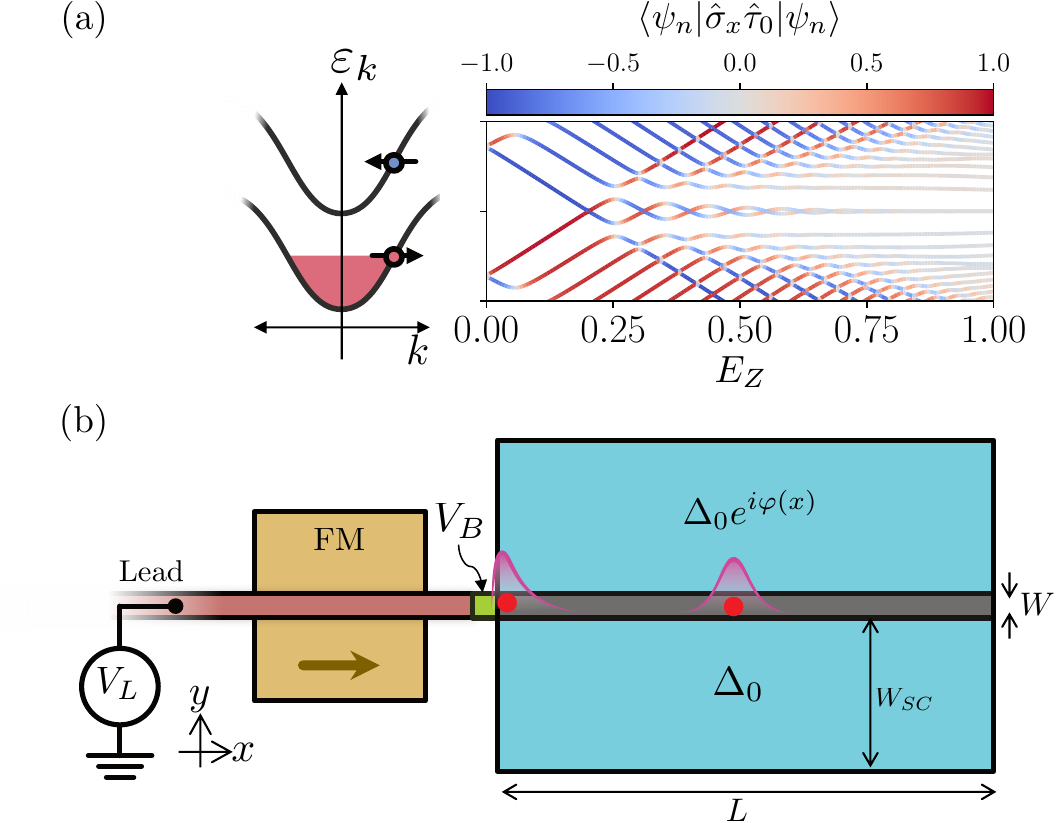}
    \caption{\textbf{(a)} Left: Band structure of electrons in spin-polarised metal lead. Right: Spectrum of the Josephson junction with the eigenstates colour-coded according to their $\hat{\sigma}_{x}\hat{\tau}_{0}$ expectation values. \textbf{(b)} Schematic of the device used to probe the spin states of the bound states in the Josephson junction when out-of-plane flux $\Phi = h/4e$. The metal lead is proximitised to a ferromagnet (FM), with magnetisation pointing in the $x$-direction (brown arrow), which spin-polarises the electron bands (top left). The lead is tunnel-coupled to the Josephson junction with a potential barrier $V_{B}$. By varying the voltage bias $V_{L}$ and measuring the current to ground, the spin properties of the sub-gap states can be probed through the Andreev conductance measurements.}
    \label{fig: andreev conductance schematic}
\end{figure}
The spin characteristics of the Josephson vortex energy levels can be effectively probed by coupling a lead to one side of the junction and measuring the Andreev conductance. However, when this measurement is performed in the presence of an out-of-plane flux $\Phi_{S}$, it fails to provide information about the sub-gap modes, as these modes are exponentially localised near the centre of the junction, as shown in Figure \ref{fig:phase diagram and wavefunctions_3}(b)(Top). To obtain a non-zero Andreev conductance at least one of the Majorana zero modes must be located at the edge of the junction. This can be achieved by tuning the out-of-plane flux to $0.5\Phi_{S}$ (a `half-vortex'), which generates a single topological domain wall in the junction, hosting a Majorana zero mode at one end \cite{stern2019fractional}. As the in-plane Zeeman coupling is reduced, the topological domain wall shifts towards the edge, resulting in the hybridisation of the two Majoranas. In the limit of large $E_{Z}$, the two Majorana zero modes become well separated, and coupling a lead to one end of the junction will result in a Andreev reflection with unit probability at zero voltage bias, due to the presence of unpaired Majorana zero mode at the end \cite{law2009majorana,lin2012zero, he2014selective}.\\ \linebreak  
Figure \ref{fig: andreev conductance schematic}(a) shows the quasi-particle energy levels colour-coded with the expectation value of the operator $\hat{\sigma}_{x}\hat{\tau}_{0}$. As $E_{Z}$ is reduced, the Majoranas hybridise to form spinful quasiparticle states. Consequently, the Andreev reflection probability becomes highly spin-dependent. This can be tested using the setup shown in Figure \ref{fig: andreev conductance schematic}(b), where a metallic lead, spin polarized in the $x$-direction by an underlying ferromagnet, is tunnel-coupled to one end of the Josephson junction. By varying the voltage $V_{L}$ and measuring the current in the wire, one can obtain the Andreev conductance as a function of $V_{L}$ for different in-plane Zeeman fields. It is expected that the Andreev conductance will exhibit a strong dependence on the sign of the spin splitting for low $E_{Z}$.\\ \linebreak 
The spin-split metallic lead is modelled by the Hamiltonian 
\begin{equation}
\mathcal{H}_{lead} = \left (\frac{-\bm{\nabla}^{2}}{2m} - \mu \right)\hat{\sigma}_{0}\hat{\tau}_{z} + E_{Z,L}\hat{\sigma}_{x}\hat{\tau}_{0},
\end{equation}
where $E_{Z, L}$ is the Zeeman splitting in the lead. A potential barrier of height $V_{B}$ is included at the interface between the semi-infinite lead and the Josephson junction. We use the Kwant toolbox to evaluate the scattering matrix of the lead-junction system at energy $eV_{L}$ \cite{groth2014kwant}:
\begin{equation}
    \hat{s}(eV_{L})= \begin{pmatrix} \hat{r}_{ee}(eV_L) & \hat{r}_{eh}(eV_L) \\ \hat{r}_{he}(eV_L) & \hat{r}_{hh}(eV_L)\end{pmatrix},
\end{equation}
where the diagonal blocks correspond to normal reflection amplitudes for electrons and holes of the potential barrier, and the off-diagonal blocks correspond to Andreev reflection amplitudes. The Andreev conductance $G_{A}(eV_{L})$ is computed using the trace formula \cite{beenakker2011random}
\begin{equation}
G_{A}(eV_L)= \frac{e^{2}}{h} \left[ N_{e} - Tr\{ \hat{r}_{ee
}^{\dagger}\hat{r}_{ee} \} +  Tr\{ \hat{r}_{he
}^{\dagger}\hat{r}_{he} \}\right],
\end{equation}
where $N_{e}$ is the number of occupied electronic bands at voltage bias $V_{L}$.\\  \linebreak 
\begin{figure}
    \centering
    \includegraphics[width=0.95\linewidth]{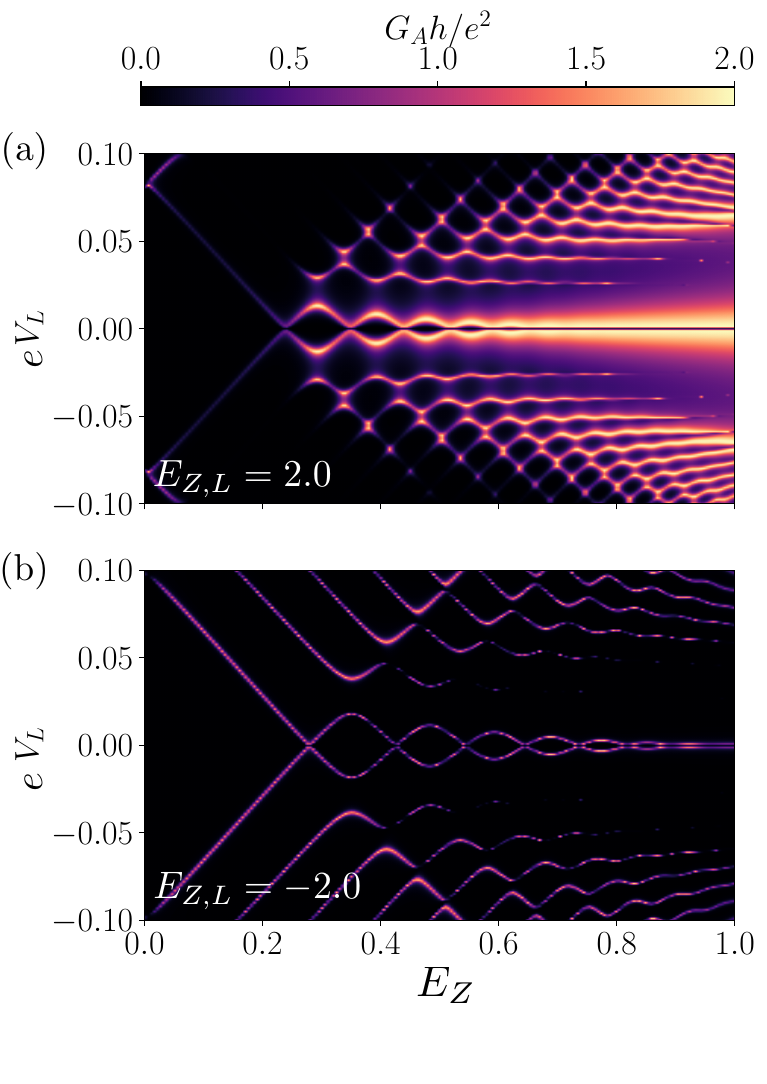}
    \caption{Andreev conductance $G_{A}$ in units of $e^{2}/h$ as a function of bias voltage $V_{L}$ and in-plane Zeeman coupling $E_{Z}$. The conductance is shown for two spin-polarising fields: (a) $E_{Z,L} = 2.0$ and (b)$E_{Z,L} = -2.0$. The potential barrier between the lead and the junction is taken to be $V_{B}$.}
    \label{fig: andreev conductance figure}
\end{figure}
As shown in Figure \ref{fig: andreev conductance figure}, the Andreev conductance depends strongly on the orientation of the spin polarisation of the metallic lead, controlled by the sign of $E_{Z,L}$. The spin splitting, $E_{Z,L}$, is chosen so that the incident electrons are fully spin-polarised for the range of voltages shown and $V_{B} = 2.0$. In Figure \ref{fig: andreev conductance figure}(a), we set $E_{Z,L} = 2.0$ which polarises all electron spins to be in the $\ket{\leftarrow}$ state, whilst in Figure \ref{fig: andreev conductance figure}(b) $E_{Z,L}= -2.0$  and all spins are in the $\ket{\rightarrow}$ state.  
We can gain a phenomenological understanding of these differences using the tunnelling characteristics of two Majorana zero modes \cite{flensberg2010tunneling}
\begin{equation}
G_{A}(eV_{L}) = \frac{2e^{2}}{h} \frac{(2eV_{L}\Gamma)^{2}}{(e^{2}V^{2}_{L} - 4\delta^{2})^{2} + (eV_{L}\Gamma)^{2}},
\end{equation}
where $\Gamma$ is the tunnel coupling between the lead and the junction, and $\delta$ is the tunnel coupling between the Majorana zero modes within the junction. the coupling $\Gamma$ is expected to decrease as the potential barrier $V_{B}$ is increased, while $\delta$ is expected to scale as $\exp(-E_{Z}/\xi)$ as discussed in Section \ref{section: Spin characteristics}. In the limit of $E_{Z}\gg \xi$ we can take $\delta \to 0$, resulting in a conductance resonance at $eV_{L} = 0$ with peak height of $2e^{2}/h$, as observed in both Figures \ref{fig: andreev conductance figure}(a) and (b). However, the difference in the width of the zero-bias resonances indicates that the $\ket{\leftarrow}$ electrons undergo stronger Andreev reflection than $\ket{\rightarrow}$ electrons, suggesting that the isolated Majorana zero mode at the edge is a coherent superposition of $\ket{\leftarrow}$ species only: $\hat{\gamma}(x) \sim \int dx f(x)\left( \hat{c}_{\leftarrow}(x) + \hat{c}^{\dagger}_{\leftarrow}(x)\right)$.  In the limit of weak $E_{Z}$, the tunnel coupling between MZMs increases, leading to two resonances at $eV_{L} = \pm \delta$. In this regime, we observe negligible conductance for $\ket{\leftarrow}$ electrons (Figure \ref{fig: andreev conductance figure}(a)) and a pair of sharp resonances for $\ket{\rightarrow}$ electrons (Figure \ref{fig: andreev conductance figure}(a)). This suggests that $\ket{\leftarrow}$ electrons undergo \textit{complete} normal reflection whilst $\ket{\rightarrow}$ electrons undergo complete Andreev reflection. This behaviour demonstrates that the resulting quasiparticle has a definite spin-polarisation in the $x$-direction. 

%% file: sections/section06.tex
\section{Conclusions and Outlook}
In this paper, we explored the spin properties and tunable coupling of Majorana zero modes in a planar Josephson junction with strong Rashba spin-orbit coupling. Our aim was to find experimentally measurable quantities that can conclusively determine the non-Abelian fusion outcome of Majorana zero modes. Our analysis reveals that the in-plane Zeeman coupling serves as a powerful tool to control the separation between Majorana zero modes, which in turn affects their coupling and the resulting energy splitting. We demonstrated that the magnetisation of the Andreev bound states, which depends on the parity of the Majorana zero modes, offers a concrete physical observable to probe these spin-dependent properties. In principle, the magnetisation may be probed by coupling the Josephson junction to a quantum dot with spin-polarised levels or an STM tip \cite{danilenko2023spin, elzerman2004single, nakajima2019quantum, bode2003spin}. 

We derived an effective Hamiltonian that accurately describes the localisation of Majorana zero modes near the center of topological domain walls. The wavefunction overlap between these modes leads to an exponentially small energy splitting, which is strongly dependent on the in-plane Zeeman field. This tunable splitting allows for a precise control of the hybridisation between Majorana modes, and thus their fusion process, thereby enabling potential applications in topological quantum computation.

Furthermore, we investigated the spin-dependent Andreev conductance as a method to probe the spin states of the sub-gap Majorana modes in the Josephson junction. By coupling a spin-polarised metallic lead to the junction, we showed that the Andreev conductance is highly sensitive to the spin polarisation of the incident electrons, especially at low in-plane Zeeman fields. This spin dependence provides a viable approach for experimentally detecting the spin characteristics of Majorana modes and their associated parity states resulting from their fusion.

In summary, our work provides a detailed understanding of the interplay between spin, parity, and coupling of Majorana zero modes in a planar Josephson junction. The insights gained from our theoretical analysis offer promising avenues for the design of spin-sensitive devices that can detect the fusion outcome of Majorana zero modes, which are critical for advancing the field of topological quantum computing.

%% file: sections/acknowledgements.tex
\section*{Acknowledgements} \label{sec:acknowledgements}
J. K. Pachos acknowledges funding from EPSRC with Grant No. EP/R020612/1. K. Ganesh and D. K.K. Lee acknowledge funding from EPSRC with Grant No. EP/R513052/1 and Grant No. EP/T51780X/1. Discussions with C. Benjamin are gratefully acknowledged. 

%% file: sections/appendix1.tex
\subsection{Tight Binding Hamiltonian }\label{sec: tight binding Hamiltonian}
For our numerical simulations we use the following tight-binding Hamiltonian:
\begin{align}\label{eqn: tight binding Hamiltonian}
    \nonumber \mathcal{H} = &\sum_{\vec{r}} \big[\left(4t - \mu \right)\sigma_{0}\tau_{z} +\Delta(\bm{r})\sigma_{0}\tau_{+} + \Delta^{*}(\bm{r})\sigma_{0}\tau_{-} \\ \nonumber &+ E_{Z}(\bm{r})\sigma_{x}\tau_{0}\big]\otimes \ket{\bm{r}}\bra{\bm{r}} \\ \nonumber 
    &+\sum_{\bm{r} , i = \{x , y\}}\left(-t\sigma_{0}\tau_{z}\otimes\ket{\bm{r}}\bra{\bm{r} + \hat{e}_{i}} + h.c.\right) \\ \nonumber &+  \frac{\alpha}{2a}\sum_{\bm{r}} \big(i \sigma_{y}\tau_{z}\otimes\ket{\bm{r}}\bra{\bm{r} + \hat{e}_{x}}  \\ &- i \sigma_{x}\tau_{z}\otimes\ket{\bm{r}}\bra{\bm{r} + \hat{e}_{y}} +  h.c. \big)
\end{align}
where $\bm{r}$ labels lattice sites, $t = 1/2ma^{2}$ and $a$ is the lattice constant of the simulation. The tight-binding calculations were performed using the Kwant library \cite{groth2014kwant}.
\subsection{Deriving the low-energy effective model} \label{sec:Andreev Hamiltonian}
In this Appendix we outline the derivation of the low-energy effective model for the Andreev bound states in the junction, and the various approximations used to get there. This gives us access to the explicit form of the Majorana zero mode spinors localised to the topological domain walls. In what follows, we work in the short junction regime where the width of the junction $W$ is much smaller than the superconducting coherence length $\xi = v_{F}/\Delta_{0}$. We will also work in the limit where the spin-orbit momentum shift is much smaller than the Fermi momentum: $m\alpha \ll k_{F}$.\\
\subsubsection{Andreev Hamiltonian}
As it stands, the Bogoliubov de Gennes Hamiltonian in Equation \ref{eqn: Bogoliubov de Gennes Hamiltonian for system} is a complicated differential operator to solve in all generality. To make progress, we will work in the Andreev approximation, which assumes that the coherence length $\xi$ is much larger than the Fermi wavelength $k_{F}^{-1}$. In this limit, eigenspinors will generally have a rapid oscillatory factor $e^{i\bm{k_{F}}\cdot\bm{r}}$, with smooth envelope functions $\vec{u}(\bm{r})$, $\vec{v}(\bm{r})$ which we wish to calculate. We will also assume the magnetic length $\ell_{B} \gg k_{F}^{-1}$. One can obtain an `Andreev Hamiltonian' for the envelope functions $\vec{u}(\bm{r})$ and $\vec{v}(\bm{r})$ by Taylor expanding the Hamiltonian around the Fermi points $\left(0 , \pm k_{F}\right)^{T}$, whilst dropping all terms $\mathcal{O}(\partial^{2}_{y})$, $\mathcal{O}(k_{x}^2)$ and higher. This gives us 
\begin{align}
\nonumber \mathcal{H}_{\pm k_{F}}(k_{x}) &\approx -i\left(\pm v_{F}\sigma_{0} - \alpha \sigma_{x}\right)\tau_{z}\partial_{y} \\ \nonumber &\mp \alpha k_{F} \sigma_{x}\tau_{z} +\alpha\sigma_{y}\tau_{z}k_{x} \\ &+ E_{Z}(y)\sigma_{x}\tau_{0} + \Delta(y) \sigma_{0}\tau_{+} + \Delta^{*}(y)\sigma_{0}\tau_{-},   
\end{align}
where the momentum operator $-i\partial_{y}:= \hat{k}_{y} \mp k_{F}$. We take the $x$-axis to be parallel to the junction and the $y$-axis to lie perpendicular to the junction of width $W$, as shown in Figure \ref{fig:Rashba_junction}. $\mathcal{H}_{\pm k_{F}}(k_{x})$ is easier to solve since it is linear in spatial derivatives. Particle-hole symmetry for the Andreev Hamiltonian is defined as 
\begin{equation}\label{eqn: particle hole symmetry Andreev hamiltonian}
    P \mathcal{H}_{k_{F}}(k_{x}) P^{-1} = -\mathcal{H}_{-k_{F}}(-k_{x}),
\end{equation}
where $P = \sigma_{y}\tau_{y} K$.
\subsubsection{Topological Phase Transitions}\label{appendix: topological phase transitions}
In the continuum, the $\mathbb{Z}_{2}$ topological index can only change when there are gap-closings at $k_{x} = 0$ \cite{kitaev2001unpaired, tewari2012topological}. In this section, we solve $\mathcal{H}(k_{x} = 0)$ for sub-gap energies $\varepsilon < \Delta_{0}$, and thus obtain the gap-closing points as a function of the Josephson phase difference $\varphi$. For clarity, we outline solutions in the vicinity of the Fermi point $(0 , k_{F})^{T}$, since solutions around $(0 , -k_{F})^{T}$ are readily obtained using the particle-hole operation defined in Equation \ref{eqn: particle hole symmetry Andreev hamiltonian}. In order to obtain the sub-gap spectrum, we must solve $\mathcal{H}_{k_{F}}(k_{x} = 0)\vec{\psi}(y) = \varepsilon \vec{\psi}(y)$ in the different regions of the system: $y < 0$ , $0 \leq y\leq W$ and $y > W$ and match wave-functions at $y = 0 , W$.\\ \linebreak 
We begin by noticing the spin conservation law at $k_{x} = 0$
\begin{equation}
[\mathcal{H}_{k_{F}}(k_{x} = 0) , \sigma_{x}\tau_{0}] = 0,
\end{equation}
which allows us to label wave-functions with their $\sigma_{x}\tau_{0}$ eigenvalues $\sigma = \pm 1$. Up to a normalization constant and overall phases, we obtain the following piece-wise spinor for the sub-gap states
\begin{widetext}
\begin{equation}\label{eqn: big equation}
\vec{\psi}_{\sigma}(y) \sim e^{i \sigma m\alpha y } \begin{cases}  \begin{pmatrix} 1 \\ \sigma\end{pmatrix}\otimes\begin{pmatrix} \Delta_{0} \\ \varepsilon + iv_{F}\kappa \end{pmatrix} e^{\kappa y} & y < 0 \\ 
A_{\sigma}\begin{pmatrix}1 \\ \sigma \end{pmatrix}\otimes\begin{pmatrix} 1 \\ 0 \end{pmatrix} \exp[i\frac{(\varepsilon - \sigma E_{Z})y}{v_{F}}] + B_{\sigma} \begin{pmatrix}1 \\ \sigma \end{pmatrix}\otimes\begin{pmatrix} 0 \\ 1 \end{pmatrix} \exp[-i\frac{(\varepsilon - \sigma E_{Z})y}{v_{F}}] & 0 \leq y \leq W \\
\begin{pmatrix} 1 \\ \sigma\end{pmatrix}\otimes\begin{pmatrix} \Delta_{0}e^{i\varphi} \\ \varepsilon - iv_{F}\kappa \end{pmatrix} e^{-\kappa(y - W)} & y > W
\end{cases}
\end{equation}
where $A_{\sigma}, B_{\sigma}$ are complex amplitudes to be determined by wave-function matching and $\kappa = \frac{\sqrt{\Delta_{0}^{2} - \varepsilon^{2}}}{v_{F}}$ is an energy-dependent decay constant for the  quasiparticle wavefunction in the superconducting region.
\end{widetext}
Matching wavefunctions at $y = 0, W$ determines the relative phase between electrons and holes, $A_{\sigma}/B_{\sigma}$, which results in the following transcendental equation for the Andreev bound state energies 
\begin{equation}
    \frac{2W}{v_{F}}\left(\varepsilon_{\sigma} - \sigma E_{Z}\right) = \varphi + 2n\pi + 2\cos^{-1}\left( \frac{\varepsilon_{\sigma}}{\Delta_{0}} \right).
\end{equation}
In the short junction limit, $v_{F}/W \gg \Delta_{0}$, we obtain the Andreev spectrum 
\begin{equation}
    \varepsilon_{\sigma}(\varphi) = \Delta_{0} \cos \left( \frac{\varphi}{2} + \frac{\sigma E_{Z} W}{v_{F}} \right),
\end{equation}
which provides an explicit derivation for the result reported in \cite{pientka2017topological}. Gap closings occur at phase differences
\begin{equation}\label{eqn: phase}
    \varphi^{k_{F}}_{\sigma} = \pi - \frac{2\sigma E_{Z} W}{v_{F}},
\end{equation}
which correspond to topological phase transitions where the $\mathbb{Z}_{2}$ index changes sign. Note we have now incorporated an extra label `$k_{F}$' in order to specify that this solution is obtained in the vicinity of the $(0 , k_{F})^{T}$ Fermi point. We also obtain a branch of solutions in the vicinity of $(0 , -k_{F})^{T}$ with gap-closings at phase differences
\begin{equation}
    \varphi^{-k_{F}}_{\sigma} = \pi + \frac{2\sigma E_{Z} W}{v_{F}}.
\end{equation}
\subsubsection{Majorana spinors}
Due to particle-hole symmetry, there are a pair of zero-energy states at each gap-closing point since the spinors $\vec{\psi}^{k_{F}}_{\sigma}(y)$ and $P\vec{\psi}^{k_{F}}_{\sigma}(y)$ are satisfy.
\begin{align}
    \mathcal{H}_{k_{F}}(0)\vec{\psi}^{k_{F}}_{\sigma}(y) &= 0  \\ \mathcal{H}_{-k_{F}}(0)P\vec{\psi}^{k_{F}}_{\sigma}(y) &= 0
\end{align}
We define the Majorana spinor basis in this degenerate manifold using the transformation 
\begin{align}
\vec{\Gamma}_{1 \sigma}(y) &= \frac{1}{\sqrt{2}}\left( \vec{\psi}^{k_{F}}_{\sigma}(y) + P\vec{\psi}^{k_{F}}_{\sigma}(y) \right) \\ 
\vec{\Gamma}_{2 \sigma}(y) &= \frac{i}{\sqrt{2}}\left( \vec{\psi}^{k_{F}}_{\sigma}(y) - P\vec{\psi}^{k_{F}}_{\sigma}(y) \right)
\end{align}
Furthermore, $\vec{\Gamma}_{i\sigma}(y)$ should be eigenvectors of the particle-hole operator. Therefore we define the following orthonormal basis kets
\begin{align}
    \ket{\chi_{\sigma}} &= \frac{1}{2}\exp\left(\frac{i\pi\sigma}{4}\hat{\sigma}_{z}\otimes\hat{\tau}_{0}\right)\begin{pmatrix}1 \\ \sigma \\ \sigma \\ -1 \end{pmatrix} \\
    \ket{\eta_{\sigma}} &= \frac{1}{2}\exp\left(\frac{i\pi\sigma}{4}\hat{\sigma}_{z}\otimes\hat{\tau}_{0}\right)\begin{pmatrix}\sigma \\ -1 \\ -1 \\ -\sigma \end{pmatrix}
\end{align}
which satisfy $P\ket{\chi_{\sigma}} = \ket{\chi_{\sigma}}$ and $P\ket{\eta_{\sigma}} = \ket{\eta_{\sigma}}$. In this basis, the Majorana spinors read 
\begin{widetext}
\begin{equation}
    \vec{\Gamma}_{1\sigma}(y) \sim \begin{cases} \big[\cos(\theta_{\alpha} y) \ket{\chi_{\sigma}} + \sin(\theta_{\alpha} y) \ket{\eta_{\sigma}}\big]e^{\Delta_{0}y / v_{F}} & y < 0 \\ \cos\left(\theta_{Z}y\right)\big[\cos(\theta_{\alpha} y) \ket{\chi_{\sigma}} + \sin(\theta_{\alpha} y) \ket{\eta_{\sigma}}\big] \\  \ \ \ \ +  \sigma \sin\left(\theta_{Z}y\right)\big[\cos(\theta_{\alpha}y)\ket{\chi_{-\sigma}} -\sin(\theta_{\alpha}y)\ket{\eta_{-\sigma}} \big] & 0 \leq y \leq W \\ \cos\left(\theta_{Z}W\right)\big[\cos(\theta_{\alpha} y) \ket{\chi_{\sigma}} + \sin(\theta_{\alpha} y) \ket{\eta_{\sigma}}\big]e^{-\Delta_{0}(y - W) / v_{F}} \\ \ \ \ \ + \sigma \sin\left(\theta_{Z}W\right)\big[\cos(\theta_{\alpha}y)\ket{\chi_{-\sigma}} -\sin(\theta_{\alpha}y)\ket{\eta_{-\sigma}} \big]e^{-\Delta_{0}(y - W) / v_{F}} & y > W
    \end{cases}
\end{equation}
\begin{equation}
    \vec{\Gamma}_{2\sigma}(y) \sim \begin{cases} \big[-\sin(\theta_{\alpha} y) \ket{\chi_{\sigma}} + \cos(\theta_{\alpha} y) \ket{\eta_{\sigma}}\big]e^{\Delta_{0}y / v_{F}} & y < 0 \\ 
    \cos\left(\theta_{Z}y\right)\big[-\sin(\theta_{\alpha} y) \ket{\chi_{\sigma}} + \cos(\theta_{\alpha} y) \ket{\eta_{\sigma}}\big] \\ \ \ \ \ - \sigma \sin\left(\theta_{Z}y\right)\big[\sin(\theta_{\alpha}y)\ket{\chi_{-\sigma}} +\cos(\theta_{\alpha}y)\ket{\eta_{-\sigma}} \big] & 0 \leq y \leq W \\ 
    \cos\left(\theta_{Z}W\right)\big[-\sin(\theta_{\alpha} y) \ket{\chi_{\sigma}} + \cos(\theta_{\alpha} y) \ket{\eta_{\sigma}}\big]e^{-\Delta_{0}(y - W) / v_{F}} \\ \ \ \ \ - \sigma \sin\left(\theta_{Z}W\right)\big[\sin(\theta_{\alpha}y)\ket{\chi_{-\sigma}} +\cos(\theta_{\alpha}y)\ket{\eta_{-\sigma}} \big]e^{-\Delta_{0}(y - W) / v_{F}} & y > W
    \end{cases}
\end{equation}
where $\theta_{\alpha} = \sigma m \alpha$ and $\theta_{Z}= \sigma E_{Z}/v_{F}$. The normalization constant for these spinors is $N = \left( v_{F}/\Delta_{0} + W\right)^{-1/2}$. These are real superpositions of the basis kets $\ket{\chi_{\sigma}}$ and $\ket{\eta_{\sigma}}$ so we see that they are themselves Majorana modes.
\end{widetext}
\subsubsection{Degenerate perturbation theory}\label{appendix: low energy effective Hamiltonian}
Therefore, the most general Majorana wavefunction is given by 
\begin{equation}
    \vec{\Gamma}(y) = \sum_{\sigma = \pm} \gamma_{1\sigma} \vec{\Gamma}_{1\sigma}(y) + \gamma_{2\sigma} \vec{\Gamma}_{2\sigma}(y)
\end{equation}
where the amplitudes $\gamma_{i\sigma} \in \mathbb{R}$. If we now switch on a small $k_{x}$ component, these amplitudes will inherit a dependence on the $x$ coordinate. Likewise, perturbing the Josephson phase away from $\varphi_{\pm}$ may cause a coupling between the Majorana spinors $\vec{\Gamma}_{1/2,\sigma}(y)$. We treat such effects using the perturbation Hamiltonian
\begin{equation}
    \delta \mathcal{H} = \delta \mathcal{H}_{k_{x}} + \delta \mathcal{H}_{\varphi},
\end{equation}
where $\delta\mathcal{H}_{k_{x}} = \alpha k_{x}\sigma_{y}\tau_{z}$ and $\delta \mathcal{H}_{\varphi} =\Delta_{0}(\varphi - \varphi_{\pm})\sigma_{0}\tau_{y} \vartheta(y - W)$. From this, a low energy effective Hamiltonian for the amplitudes $\gamma_{i\sigma}(x)$ may be obtained by projecting $\delta \mathcal{H}$ onto the basis $\vec{\Gamma}_{i\sigma}(y)$ \cite{fu2008superconducting}: 
\begin{equation}
    H_{eff} = \int dx \sum_{ij,\sigma\sigma'} \gamma_{i\sigma}(x)\bra{\Gamma_{i\sigma}}\delta \mathcal{H}\ket{\Gamma_{j\sigma'}}\gamma_{j\sigma'}(x).
\end{equation}
To this end, it is useful to write down the matrix elements of $\sigma_{y}\tau_{z}$ in the basis $\{ \ket{\chi_{\pm}} , \ket{\eta_{\pm}}\}$: 
\begin{align}
\bra{\chi_{\sigma}}\sigma_{y}\tau_{z}\ket{\chi_{\sigma'}} &= -\delta_{\sigma\sigma'} \\ \bra{\eta_{\sigma}}\sigma_{y}\tau_{z}\ket{\eta_{\sigma'}} &= +\delta_{\sigma\sigma'} \\ \bra{\chi_{\sigma}}\sigma_{y}\tau_{z}\ket{\eta_{\sigma'}} &= 0
\end{align}
Projecting $\delta \mathcal{H}_{k_{x}}$ onto the basis $\vec{\Gamma}_{i\sigma}(y)$ we get 
\begin{equation}
    [\delta\mathcal{H}^{ij}_{k_{x}}] = \tilde{v}k_{x} \begin{pmatrix} -\cos(\theta_{\alpha}W) & \sin(\theta_{\alpha}W) \\ \sin(\theta_{\alpha}W) & \cos(\theta_{\alpha}W)\end{pmatrix},
\end{equation}
where we have defined the effective group velocity
\begin{equation}
\tilde{v}  \approx \frac{\alpha \Delta^{2}_{0}[ mv_{F}\alpha\cos(m\alpha W) + \Delta_{0}\sin(m\alpha W)]}{m\alpha v_{F}((m\alpha v_{F})^{2} + \Delta^{2}_{0})}.
\end{equation}
In the above expression we have dropped terms $\mathcal{O}(\Delta_{0}W/v_{F})$. Projecting $\delta \mathcal{H}_{\varphi}$ onto the basis $\vec{\Gamma}_{i\sigma}(y)$ we get  
\begin{equation}
    [\delta\mathcal{H}^{ij}_{\varphi}] = \frac{\Delta_{0}}{2}(\varphi - \varphi_{\sigma})\cos\left(\frac{2WE_{Z}}{v_{F}}\right)\begin{pmatrix}0 & -i \\ i & 0 \end{pmatrix}.
\end{equation}
Finally we perform the following SO(2) basis rotation on $\left(\gamma_{1\sigma} , \gamma_{2\sigma}\right)^{T}$ 
\begin{equation}
    \begin{pmatrix}\tilde{\gamma}_{1\sigma} \\ \tilde{\gamma}_{2\sigma} \end{pmatrix} \to 
    \begin{pmatrix} \sin(\theta_{\alpha}W/2) & \cos(\theta_{\alpha}W/2) \\  \cos(\theta_{\alpha}W/2) & -\sin(\theta_{\alpha}W/2)\end{pmatrix}\begin{pmatrix}\gamma_{1\sigma} \\ \gamma_{2\sigma} \end{pmatrix}
\end{equation}
which gives us the effective Hamiltonian for each spin sector $\sigma = \pm$
\begin{equation}\label{eqn: effective Hamiltonian}
\mathcal{H}^{\sigma}_{eff} = -i\tilde{v}\partial_{x}\hat{\nu}_{z} + \frac{\Delta_{0}}{2}(\varphi - \varphi_{\sigma})\cos\left(\frac{2WE_{Z}}{v_{F}}\right) \hat{\nu}_{y}.
\end{equation}
The Pauli matrices $\hat{\nu}_{i}$ act on the basis $\left(\tilde{\gamma}_{1\sigma} , \tilde{\gamma}_{2\sigma}\right)^{T}$. Equation \ref{eqn: effective Hamiltonian} gives us an effective description for Majorana modes in the junction for a  constant superconducting phase difference $\varphi$ across the junction near  the boundary
of the topological phase transition  (diamond in Figure \ref{fig:phase diagram and wavefunctions_3}(a)). 

%% file: main.bbl
\begin{thebibliography}{10}
\bibitem{vecpotentialandscreening}
We have used a vector potential gauge where $\bm{A} = B_{z}y\hat{x}$ for $0 \leq y \leq W$, $\bm{A} = \bm{0}$ for $y < 0$ and $\bm{A} = B_{z}W \hat{x}$ for $y > W$. The linear phase texture is valid if $L$ is much shorter than the Josephson penetration depth $\lambda_{J}$ because we can neglect the magnetic field produced by screening currents.
\end{thebibliography}
